# Shaping the Epochal Individuality and Generality: The Temporal Dynamics of Uncertainty and Prediction Error in Musical Improvisation

Tatsuya Daikoku [1,2,3]
[1] Graduate School of Information Science and Technology, The University of Tokyo
[2] Centre for Neuroscience in Education, University of Cambridge
[3] Center for Brain, Mind and KANSEI Sciences Research, Hiroshima University

**\* Corresponding author**
Tatsuya Daikoku
Graduate School of Information Science and Technology, The University of Tokyo, 7-3-1 Hongo, Bunkyo-ku, Tokyo, 113-8656, Japan
Email: daikoku.tatsuya@mail.u-tokyo.ac.jp

## Competing Interests

The authors declare no competing financial interests.

## Author Contributions

T.D. conceived the method of data analyses, and collected the data. T.D, analyzed the data, prepared the figures, wrote the manuscript. and finalized the manuscript.

## Acknowledgements

This research was supported by JSPS KAKENHI (22KK0157; 22H05210; 21H05063) and JST Moonshot Goal 9 (JPMJMS2296), Japan. The funding sources had no role in the decision to publish or prepare the manuscript.

## Data Availability

The scripts for the computational model (Hierarchical Bayesian Statistical Learning: HBSL) and analysis and all data including results have been deposited to an external source (https://osf.io/xj6kf/?view_only=1c2f946e7d994e61abc23e311f9b8e04). The other data and results of statistical analysis are shown in supplementary data.

# Abstract

Human improvisational acts contain an innate individuality, derived from one's experiences based on epochal and cultural backgrounds. Musical improvisation, much like spontaneous speech, reveals intricate facets of the improviser's state of mind and emotional character. However, the specific musical components that reveal such individuality remain largely unexplored. Within the framework of brain's statistical learning and predictive processing, this study examined the temporal dynamics of uncertainty and surprise (prediction error) in a piece of musical improvisation. This cognitive process reconciles the raw auditory cues, such as melody and rhythm, with the brain's internal models shaped by its prior experiences.

This study employed the Hierarchical Bayesian Statistical Learning (HBSL) model to analyze a corpus of 456 Jazz improvisations, spanning 1905 to 2009, from 78 distinct Jazz musicians. The results indicated distinctive temporal patterns of surprise and uncertainty, especially in pitch and pitch-rhythm sequences, revealing era-specific features from the early 20th to the 21st centuries. Conversely, rhythm sequences exhibited a consistent degree of uncertainty across eras. Further, the acoustic properties remain unchanged across different periods.

These findings highlight the importance of how temporal dynamics of surprise and uncertainty in improvisational music change over periods, profoundly influencing the distinctive methodologies artists adopt for improvisation in each era. Further, it is suggested that the development of improvisational music can be attributed to the brain's adaptive statistical learning mechanisms, which constantly refine internal models to mirror the cultural and emotional nuances of their respective epochs. This study unravels the evolutionary trajectory of improvisational music and highlights the nuanced shifts artists employ to resonate with the cultural and emotional landscapes of their times. Such shifts in improvisational ways offer a window into understanding how artists intuitively respond and adapt their craft to resonate with the cultural zeitgeist and the emotional landscapes of their respective times.

# 1. Introduction

## 1.1. Statistical Learning and Emergence of Individuality

Human improvisational acts arise subconsciously, interwoven with emotion [1,2]. Especially, auditory improvisation such as musical jazz and spontaneous spoken utterances entails the individuality of behavioural or psychological patterns shaped by the improvisor's experiences and history [3]. For example, just as a person might shout when angered or speak in a lively manner when joyful, spontaneous verbal outputs often encapsulate her or his emotional state and character. Such individual traits of emotion and character from spontaneous speech can be predicted from acoustic properties such as spectral and temporal frequency patterns [4,5]. While musical production is also thought to be tied to the individuality of behavioural or psychological patterns, our understanding of which musical and acoustic components enable the visualization of such individuality remains limited.

Recent research increasingly explores how the brain learns and produces music, often grounding these investigations in the framework of predictive processing [6]. In musical contexts, this process seeks to reconcile bottom-up auditory cues, like melody and rhythm, with the brain's top-down predictions, which are shaped by its internal models [7,8]. In the brain's predictive processing framework, evidence has suggested the acquisition of musical knowledge is deeply rooted in statistical learning mechanisms of the brain [9,10]. This inherent learning mode, pivotal in brain development [11], shapes our perception and generation of both music and language. At its core, statistical learning computes the transition probability of sequential information and the uncertainty of its probability distribution. Human can predict a forthcoming event from a sequential information using an internal probabilistic model acquired through statistical learning.

Crucially, perceptual uncertainty and expectations are not innately tied to music itself [12]. Instead, they are molded by one's auditory experiences. For instance, someone raised within a particular era or culture might find music from an entirely different time or culture more uncertain and harder to predict each tone during hearing it. Conversely, music from one's own era and culture becomes more predictable and less uncertain. This phenomenon arises as individuals continually refine their internal models via statistical learning, crafting a music probabilistic model tailored to their specific era and culture. Both neural and computational research has indicated that the individual difference in music expectation associated with music production (i.e., composition) [13, 14] and perception [15] are influenced by past experiences in statistical learning.

Improvisational acts in music, such as those found in jazz, are also linked to the brain's predictive processing. Using a musical task that involved rating chord progressions, Przysinda et al. [16] compared predictive processing between jazz improvisers, non-improvising musicians, and non-musicians. They found that jazz musicians exhibited a preference for unpredictable (surprising) chord progressions. Moreover, these unpredictable progressions triggered more substantial music expectancy-related neural responses in jazz musicians. This implies that people who can precisely predict a musical event often favour the unpredictable, possibly due to their enhanced ability to discriminate between familiar and novel musical elements.

Further, using a computational model of the brain's statistical learning, a previous study examined the statistical characteristics of jazz improvisation played by globally renowned jazz pianists: Bill Evans, Herbie Hancock, and McCoy Tyner [17]. The analysis illuminated individual differences in statistical patterns between players, while also identifying shared characteristics among them. This underscores the dual role of statistical learning in shaping both the uniqueness and generality of improvisational music.

## 1.2. Temporal Dynamics of Uncertainty and Prediction

Within the framework of predictive processing in the brain, the insight into understanding the individual difference in musical improvisation is that it embeds both predictable and unpredictable tones throughout a music piece, forming temporal dynamics of surprise and uncertainty. This may relate to the so-called "musical form". For instance, emotional responses to prediction errors, like excitement from surprise, intensify after a sequence of expected stimuli, compared to unexpected stimuli [18]. Similarly, while a prolonged series of expected stimuli might lead to boredom and waning interest in the music, following an unexpected stimulus with an expected one can evoke a feeling of relief. Thus, the representation of musical emotion including preference can be determined by the temporal fluctuations between these predictable and unpredictable musical elements [19].

Prior research suggests that the individuality of musical form depends on historical epochs rather than composers [20, 21]. Intriguingly, even within a single composer's oeuvre, there can be variations in uncertainty based on the era of composition [22]. In essence, individuality in musical improvisation might stem from the prevailing trends of the era or shifts in a musician's personal experiences. For instance, if musicians consistently play the same pieces within a specific period, or repeatedly perform the same compositions, the inherent musical model of that time—or even within the individual—can lose its uncertainty and novelty. This reduced uncertainty can lead to a waning interest in both performance and listening, prompting a desire for different musical expressions. Such mental eagerness to generate novel music could underpin the evolving styles of music across different eras and developmental stages. While statistical characteristics of a music piece, rooted in statistical learning, have been posited to manifest differences across performers [17], it remains unclear if distinct temporal patterns of surprise and uncertainty are similarly discernible. This study examines these temporal fluctuations or dynamics of surprise and uncertainty, aiming to unravel the unique temporal patterns across different eras, individuals, cultures, musical styles, and instruments.

## 1.3. Purpose of The Present Study

The present study examined how the individual difference of musical improvisation emerges in the temporal patterns of surprise and uncertainty, using the Hierarchical Bayesian Statistical Learning (HBSL) model [23], mimicking the "hierarchical" processing of statistical learning in the brains. Recent studies have suggested two types of hierarchical statistical learning systems [24, 25]. The first system constitutes the fundamental function of statistical learning, which groups chunks of information with high transition probabilities and integrates them into a cohesive unit. The second system involves statistical learning that arranges various chunked units to form a hierarchical syntactic structure (Figure 1). Therefore, statistical learning plays a critical role in acquiring the hierarchy, a unique and essential feature of language and music [26]. The HBSL model simulates such an hierarchical process of statistical learning [23].

The HBSL model computes the Shannon information content (surprise) and entropy (uncertainty) based on transitional probabilities [27] of tone sequence from a corpus of 456 Jazz improvisation played from 1905 to 2009 years by 78 different Jazz musicians, as the training data. This study calculated the temporal dynamics of two types of values through statistical learning: Bayesian surprise (or prediction errors) and Uncertainty (or entropy) in each pitch sequence, rhythm sequence, and sequence combining pitch and rhythm (hereafter, pitch-rhythm sequence). Further, we also analyzed acoustic properties in each music, particularly the rhythm domain below 40 Hz and the pitch domain.

This study hypothesized that the unique temporal dynamics of improvisational music have gradually evolved over time. Previous research suggests that the acoustic characteristics of music, such as pitch and rhythm, remain relatively consistent across different eras, cultures, and genres [28-30]. Given this, while the temporal patterns of surprise and uncertainty may shift, the acoustic properties of improvisational performances likely remain stable across different periods.

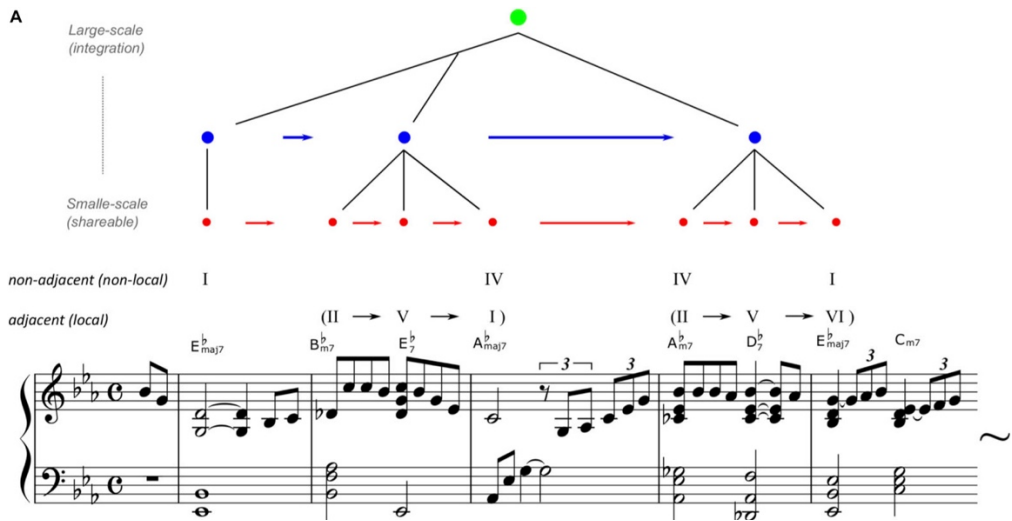

**Figure 1**. **Hierarchical statistical learning of music.** Reprinted from Daikoku et al. [25]. A segment from "Misty" by Errol Garner (1954) illustrates the principle. The displayed arrangement is simplified, highlighting only major/minor distinctions, flats, and 7th notes to emphasize the prevalent "two-five-one (II–V–I)" chord progression. For example, jazz music has general regularities in chord sequences such as the so-called "two-five-one (II–V–I) progression." It is a succession of chords whose roots descend in fifths from the supertonic (II) to dominant (V), and finally to the tonic (I). Such syntactic progression frequently occurs in jazz music, and therefore, the statistics of the sequential information have high transitional probability and low uncertainty. Thus, once a person has learned the statistical characteristics, it can be chunked as a commonly used unit among improvisers. In contrast, the ways of combining the chunked units are different between musicians.

## 2. Results

This study applied the HBSL model incorporating the Bayesian reliability of probabilities into a Markov model [18], which simulates statistical learning processes of the brain based on a series of our neurophysiological evidence (e.g., [15, 31, 32]). This model can not only calculate the transition probabilities but also determine the reliability of the transition probabilities from the inverse of the variance of the prior probability distribution. Using the normalized values of transition probabilities and reliability, this model chunks transition patterns when the product of reliability * probability is greater than a constant c. The constant can be decided based on the sample length and the number of learning trials. This study defined c=5 given the sample length used in this experiment. A chunked unit can be further integrated with another chunked unit, generating a longer unit in a higher hierarchy (e.g., from red to blue or from blue to green in Figure 1). That is, by the cascade of chunking during statistical learning, the model gradually forms the hierarchical structure. This model can derive the surprise and uncertainty of every tone in the tone sequence.

The HBSL model computes the information content and entropy based on transitional probabilities of tone sequence from a corpus of 456 Jazz improvisations. This corpus is a collection of 456 annotated improvisation recordings from well-known jazz musicians [33]. Each music was played by either of 13 different instruments including alto saxophone, bass clarinet, baritone saxophone, clarinet, cornet, guitar, piano, soprano saxophone, trombone, trumpet, tenor saxophone, C melody tenor saxophone, and vibraphone, and 8 different styles including Bebop, Cool, Free, Fusion, Hardbop, Post bop, Swing, and Traditional. This study calculated the temporal dynamics of two types of values through statistical learning: Bayesian surprise (or prediction errors) and Uncertainty (or entropy) in each pitch sequence, rhythm sequence, and sequence combining pitch and rhythm (hereafter, pitch-rhythm sequence). The Bayesian surprise was measured by the Kullback-Leibler divergence between a distribution P(x) before learning an event ($e_n$) and a distribution Q(x) after learning the event ($e_{n+1}$).

### 2.1. Probabilistic Space: Temporal Dynamics of Surprise and Uncertainty

To understand the characteristics of the temporal dynamics of both surprise and uncertainty in each of pitch sequence, rhythm sequence, and pitch-rhythm sequence, these time courses of each surprise and uncertainty were dimensionally reduced into two dimensions using t-distributed stochastic neighbor embedding (tSNE). The results revealed specific characteristics of temporal dynamics, particularly during the early part of the 20th to the 21st centuries (highlighted in black in Figure 3), especially in terms of uncertainty within pitch (a) and pitch-rhythm sequences (c). Conversely, rhythm (Figure 3, b) did not demonstrate notable characteristics of temporal dynamics across different periods. Similarly, in the surprise or probability (Figure 2), pitch-rhythm sequences (c) exhibited specific characteristics of temporal dynamics, especially during the early periods (highlighted in black). Like with uncertainty, rhythm did not display any era-specific characteristics in temporal dynamics. These findings suggest that rhythm consistently demonstrates a general degree of uncertainty and probabilistic fluctuations regardless of the era. In contrast, pitch or pitch-rhythm appears to have era-specific characteristics. Intriguingly, distinctions among players, instruments, or styles were not detected in either of pitch, rhythm, or pitch-rhythm (Supplementary, Figure A).

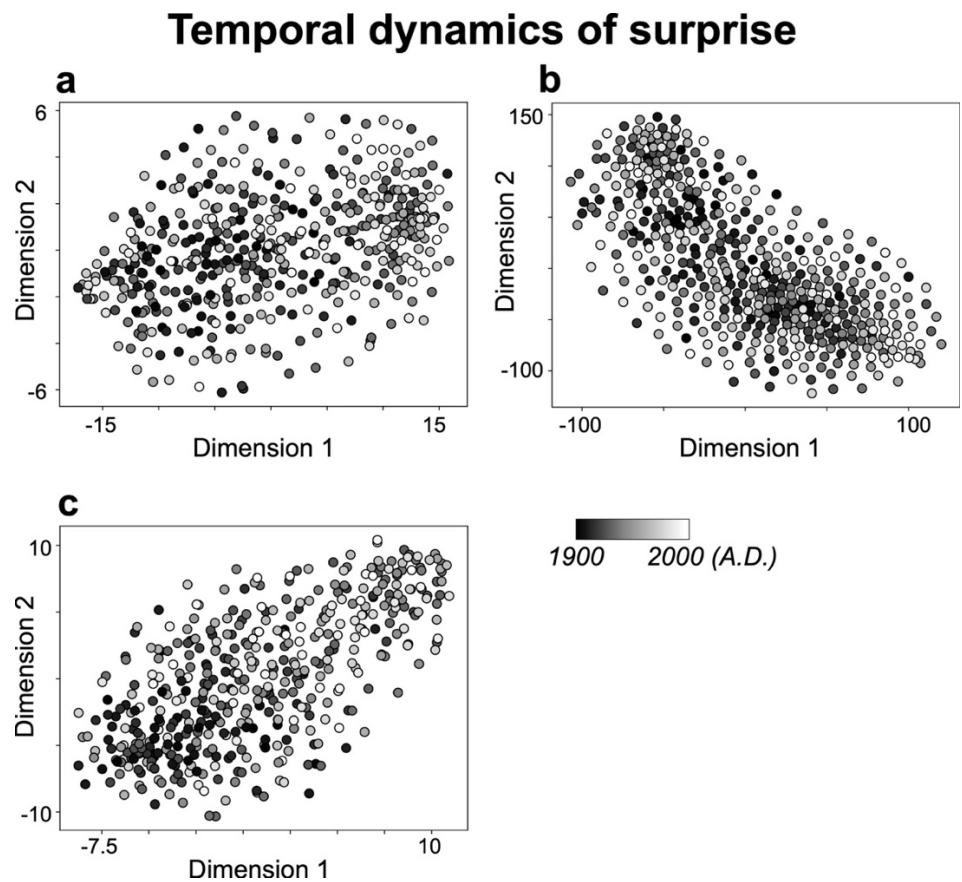

**Figure 2**. Characteristics of temporal dynamics of surprise (inverse of probability values) in pitch (a), rhythm (b), and pitch-rhythm (c) sequences, using tSNE.

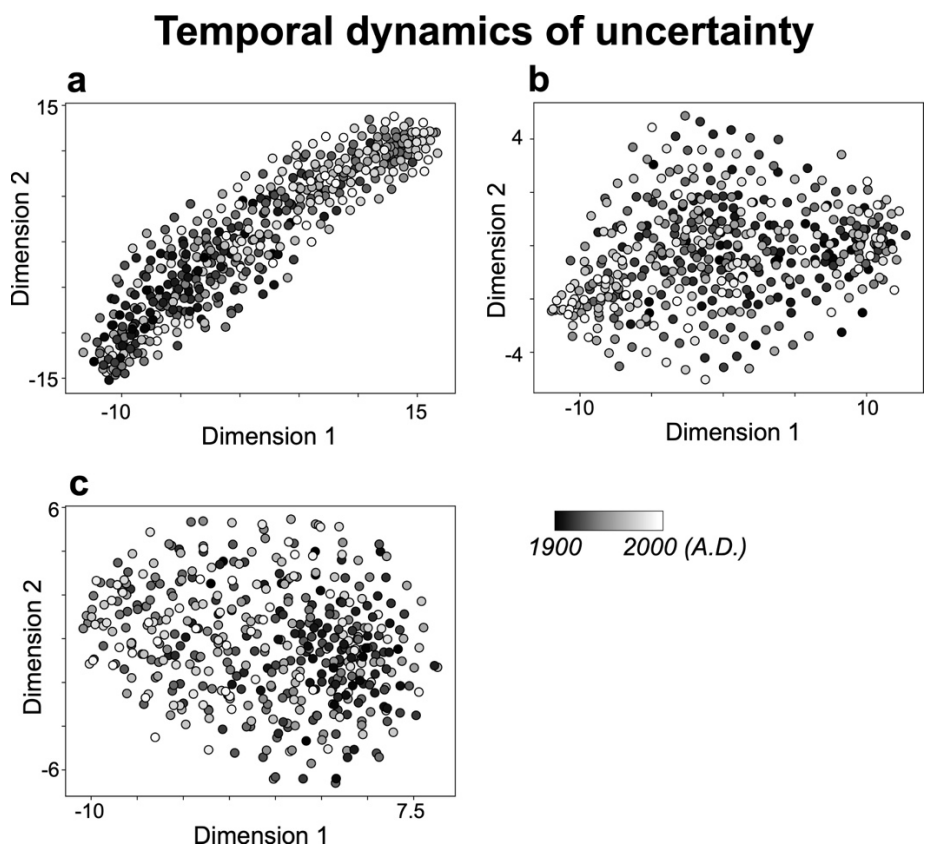

**Figure 3**. **Characteristics of temporal dynamics of uncertainty (entropy values) in pitch (a), rhythm (b), and pitch-rhythm (c) sequences, using tSNE**.

## 2.2. Acoustic Space

This study also examined acoustic characteristics in each music. All of the MIDI data were converted into WAV format to extract the rhythm waveform (modulation wave) below 40 Hz and spectral waveform (carrier waveform) using the Bayesian probabilistic amplitude modulation model (PAD, [34]). PAD utilizes Bayesian inference to estimate the most suitable modulator (i.e., envelope) and carrier that best align with the data and a priori assumptions. The resulting solution takes the form of a probability distribution, which describes the likelihood of a specific setting of modulator and carrier given the observed signal. Thus, PAD summarizes the posterior distribution by returning the specific envelope and carrier with the highest posterior probability, thereby providing the best fit to the data. In the present study, we manually entered the PAD parameters to produce the modulators at an oscillatory band level (i.e., <40 Hz) isolated from a carrier at a higher frequency rate (>40 Hz). The carrier reflects components, including noise and pitches.

In each sample, the modulators (envelopes) were converted into time-frequency domains using scalogram. The scalograms depict the amplitude modulation (AM) envelopes derived by recursive application of probabilistic amplitude demodulation. We then calculated the average frequency power at each frequency and further averaged it based on each decade played, each genre, each instrument, and each player.

The result detected that the acoustic properties of pitch and rhythm frequencies (Figure 4) showed no specific characteristics between different decades. They showed a peak power at around 500 Hz in pitch frequency and a peak power at lower frequency (i.e., slower rhythm) in rhythm frequency as demonstrated 1/f power low [35].

We also investigated the horizontal rate patterns by computing the cycle lengths of the AM waveform. First, troughs were identified because they reflected the boundaries of the edges between cycles. After detecting all troughs in the temporal rate bands, the cycle lengths were determined by calculating the length between adjacent troughs. Then we calculated the horizontal rate. For example, if the length of a cycle (AM c1) is 500 ms and the length of the subsequent cycle (AM c2) is 500 ms, then the horizontal rate is 0.5 (i.e., 1:1) based on the formula of $c_1/(c_1+c_2)$ [28]. These values were calculated for each cycle in each music sample. We then averaged the cycle lengths per music.

The findings showed that the probabilistic density for cycle rate in the modulation envelope (rhythm) also showed no specific characteristics between different decades (Figure 5). The probability densities of a 1:1 Rate was statistically stronger than the other rates. Then, the probability densities of the 1:2 and 2:1 rate was also relatively stronger than the other rates.

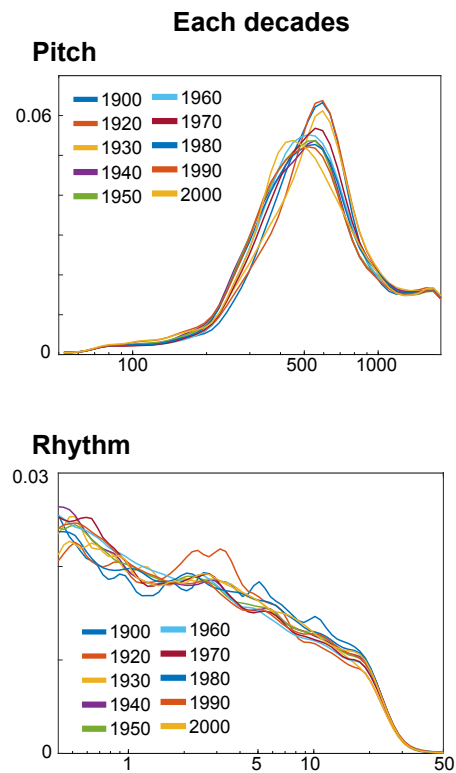

**Figure 4**. **Acoustic properties of spectral frequency (pitch) and temporal frequency (rhythm, envelope of waveform) in each decade**.

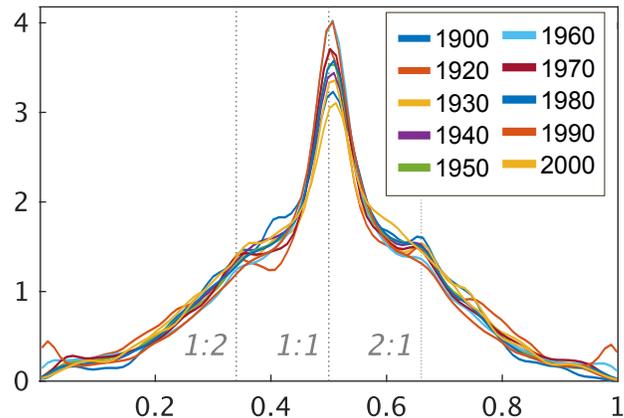

**Figure 5**. **Probabilistic density for cycle rate in the modulation envelope (rhythm) in each decade**. Based on the formula of $c_1/(c_1+c_2)$, the 0.5 (1:1) rate indicates that the relationship between the length of a given cycle (AM $c_1$) and that of the subsequent cycle (AM $c_2$) is equivalent, while a 1:2 rate refers to a situation where the length of the succeeding cycle is twice that of the preceding cycle. The probability densities of a 1:1 Rate was statistically stronger than the other rates. Then, the probability densities of the 1:2 and 2:1 rate was also relatively stronger than the other rates.

## 3. Discussion

This study investigated how individual differences in musical improvisation appear in the temporal patterns of surprise and uncertainty using the computational model of brain's statistical learning. The results suggested specific characteristics in the temporal patterns, particularly from the early 20th to the 21st centuries, with pitch and pitch-rhythm sequences embodying era-specific features.

Notably, the temporal patterns of surprise and uncertainty in rhythm sequences did not demonstrate any era-specific characteristics, suggesting a consistent degree of uncertainty and probabilistic fluctuations in the sequence throughout different periods (Figure 2b and Figure 3b). This consistency in rhythm contrasts sharply with pitch (Figure 2a and Figure 3a) and pitch-rhythm sequences (Figure 2c and Figure 3c), which were characterized by temporal fluctuations of surprise and uncertainty unique to particular epochs. Prior research has also found distinctive patterns in pitch probability distributions among individual performers in improvisation, while rhythmic distributions remained consistent across performers [17]. This hints at the possibility that while the temporal pattern of rhythm might transcend individual and temporal differences, displaying a more universal character, the temporal pattern of pitch, often considered crucial for individual emotional expression in music [36] as well as speech [37], may indeed reveal individual differences in expression [38].

On the other hand, it's worth noting that distinct differences tied to individual players, instruments, or musical styles were elusive (see Figure A in the Supplementary material). Contrary to previous research which identified musicians' uniqueness in the probability distribution without considering (episodic) time information [17], this study focused on the time information: the temporal fluctuations of probability, encompassing a mix of surprise (prediction error) and correct expectation, contributing to a narrative or episodic information. The findings in this study may imply that, particularly in the temporal pattern of probability, the individual difference in improvisation studied here predominantly arose from larger epochal shifts rather than individual contributions.

The importance of temporal information to extract individual difference in improvisation was also indicated at the acoustic level (Figure 4). That is, within the realm of acoustic properties, there wasn't a clear distinction in the frequencies of pitch and rhythm when comparing different eras. This observation lends weight to the proposed hypothesis that, even though the temporal patterns of surprise and uncertainty may undergo changes over time, the fundamental acoustic properties without time information seem to exhibit a remarkable consistency across various periods. For instance, the notable peak in power findings around 500 Hz in pitch frequency, and the lower frequency for rhythm align with the established understanding of 1/f power law [30]. These results further emphasize the idea that certain frequency characteristics remain stable and consistent, irrespective of the era and genre in which they are observed [29].

In addition to other parameters, this study delved into the examination of the horizontal ratio, a key metric that informs our understanding of rhythmic patterns. Intriguingly, a 1:1 ratio consistently emerged, signaling a widespread tendency towards equilibrium or balance in rhythmic constructs (Figure 5). Furthermore, simpler integer ratios, such as 1:2 and 2:1, were also observed. Previous research suggests that such horizontal ratios are ubiquitous across varied species, diverse cultures, and disparate musical genres [28, 30]. Such consistency strengthens the argument that there exists a set of rhythmic patterns that might be universally resonant and preferred, irrespective of cultural or biological differences.

Broadly reflecting on our findings, it appears that the evolutionary trajectory of improvisational music is intricately woven into the temporal dynamics of surprise and uncertainty. While certain elements, such as the acoustic properties, demonstrate a remarkable level of stability and consistency across periods, the essence of improvisation, which is deeply rooted in the temporal patterns of surprise and uncertainty, undergoes a profound transformation as it navigates through varying epochs. This intricate observation leads us to infer that even though the foundational "acoustic" properties of music might predominantly remain unchanged, there is a developmental change in the "probabilistic" properties, particularly from the early 20th to the 21st. This evolution is seemingly in tandem with, and perhaps a response to, the prevailing zeitgeist and predictability of each distinct era. Such a duality in musical development, we posit, might very well be attributed to the intricate workings of the brain's "adaptive" statistical learning mechanisms [32, 39, 40]. These mechanisms, in their continual quest for adaptation, incessantly refine internal probabilistic models to suit the cultural and emotional demands of their respective epochs [15].

While our study provides significant insights, it is important to acknowledge its inherent limitations. A primary concern lies in its focus solely on jazz improvisations. This emphasis might constrain the applicability of our findings to broader musical genres, potentially overlooking the nuances that might be present in other forms of improvisation or diverse musical styles like Western classical music, various ethnic music, and more. Additionally, despite analyzing an expansive corpus of compositions, our dataset predominantly spans the 20th and 21st centuries. This temporal focus leaves a considerable gap, neglecting the potentially rich and varied improvisational intricacies of music from earlier historical periods. Furthermore, the observation of consistent acoustic properties across different eras might be influenced by the specific parameters we selected for the Bayesian probabilistic amplitude modulation model. Future studies might benefit from either refining these parameters or introducing a broader spectrum of variables. This could allow for the detection of more nuanced variations and further our understanding of the evolution of improvisational techniques.

To conclude, this study examines how musical improvisation has changed over time, giving us a deeper understanding of this ever-evolving art form. Our main discovery highlights the importance of how temporal dynamics of surprise and uncertainty in improvisational music change over periods, profoundly influencing the distinctive methodologies artists adopt for improvisation in each era. These changes aren't just random;

they clearly mark different times in music history. Such shifts in improvisational techniques offer a window into understanding how artists intuitively respond and adapt their craft to resonate with the cultural zeitgeist and the emotional landscapes of their respective times. As we embark on future research endeavors in this domain, a meticulous appreciation of these subtleties becomes paramount, ensuring a holistic understanding that seamlessly merges the artistry with the underlying scientific principles of musical improvisation.

## 4. Conclusion

In conclusion, this research provides a comprehensive understanding of the characteristics of temporal dynamics of surprise and uncertainty in musical improvisation, particularly within jazz, from the 20th to the 21st centuries. Our findings illuminate the contrast between the consistency of acoustic properties across eras and the transformation of probabilistic properties. Such transformation can be attributed to the brain's adaptive statistical learning mechanisms, which adjust internal models to resonate with the emotional and cultural demands of their era. Further research could gain valuable insights by exploring other forms of improvisation or diverse musical styles like Western classical music, various ethnic music, and more. Such a study will deepen our understanding of how music has evolved over time.

# 5. Methods

## 5.1. Hierarchical Bayesian Statistical Learning Model

This study applied a neuro-inspired hierarchical Bayesian Statistical Learning (HBSL) model incorporating the Bayesian reliability of probabilities into a Markov model [18], which simulates statistical learning processes of the brain based on a series of neurophysiological evidence (e.g., [15, 31, 32]). The python codes of this model have been deposited to an external source (https://osf.io/cqmz8/?view_only=b95d94626a364700adb9e1e94384525d). This computes transitional probability from sequences [11], grasps uncertainty/entropy [41], and predicts a future state based on the internalized statistical model. Transition probability, based on Bayes' theorem, determines the likelihood of a subsequent event ($e_{n+1}$) given previous occurrence ($P(e_{n+1}|e_n)$). From a psychological perspective, the transitional probability ($P(e_{n+1}|e_n)$) can be interpreted as positing that the brain predicts a subsequent event $e_{n+1}$ based on the preceding events $e_n$ in a sequence. Psychologically, this suggests our brain anticipates an upcoming event ($e_{n+1}$) based on the most recent preceding events ($e_n$) in a certain sequence. In other words, learners expect the event with the highest transitional probability based on the latest n states, whereas they are likely to be surprised by an event with a lower transitional probability. Furthermore, transitional probabilities are often translated as information contents:

$$I(e_{n+1}) = -\log_2 P(e_n+1|e_n) \qquad (1)$$

The lower information content (i.e., higher transitional probability) means higher predictabilities and smaller surprising, whereas the higher information content (i.e., lower transitional probability) means lower predictabilities and larger surprising. In the end, a tone with lower information content may be one that a composer is more likely to predict and choose as the next event, compared to tones with higher information content. The information content can be used in computational studies of music to discuss psychological phenomena involved in prediction and statistical learning. the entropy of chord $e_{n+1}$ is the expected information content of chord $e_n$. This is obtained by multiplying the conditional probability of all possible chords in S by their information contents and then summing them together, giving:

$$H(e_{n+1}) = -\Sigma p(e_{n+1} = e|e_n)\log_2 p(e_{n+1} = e|e_n) \qquad (2)$$

Entropy gauges the perceptual uncertainty a listener feels in predicting an upcoming tone based on prior tones, while information content quantifies the surprise experienced upon hearing the actual chord. The HBSL model applies a Dirichlet distribution as a prior distribution, which can not only calculate the transition probabilities but also determine the "reliability" of the transition probabilities from the inverse of the variance of the prior probability distribution. Using the normalized values of transition probabilities and reliability, this model chunks transition patterns when the product of "reliability * probability" is greater than a constant c. The constant can be decided based on the sample length and the number of learning trial. In this study, we defided c=5 given the sample length used in this experiment. A chunked unit can be further integrated with another chunked unit, generating a longer unit in higher hierarchy (e.g., from red to blue or from blue to green in Figure 1). That is, by the cascade of chunking during statistical learning, the model gradually forms the hierarchical structure. This model can derive the surprise and uncertainty of every tone in the tone sequence.

## 5.2. Materials and Procedure

The HBSL model computes the Shannon information content and entropy based on transitional probabilities [27] of tone sequence from a corpus of 456 Jazz improvisation (The Jazzomat Research Project, https://jazzomat.hfm-weimar.de/index.html) played from 1905 to 2009 years by 78 different Jazz musicians, as the training data. For information on the music corpus, see https://osf.io/xj6kf/?view_only=1c2f946e7d994e61abc23e311f9b8e04: the Weimar Jazz Database. This corpus is a collection of 456 annotated improvisation recordings1 from well-known jazz musicians [33]. Each music was played by either of 13 different instruments including alto saxophone, bass clarinette, baritone saxophone, clarinet, cornet, guitar, piano, soprano saxophone, trombone, trumpet, tenor saxophone, C melody tenor saxophone, and vibraphone, and 8 different styles including Bebop, Cool, Free, Fusion, Hardbop, Postbop, Swing, and Traditional.

This study calculated the temporal patterns of two types of values through statistical learning: Bayesian surprise (or prediction errors) and Uncertainty (or entropy) in each pitch sequence, rhythm sequence, and pitch-rhythm sequences. The Bayesian surprise was measured by the Kullback-Leibler divergence between a distribution P(x) before learning an event ($e_n$) and a distribution Q(x) after learning the event ($e_{n+1}$) . The Kullback-Leibler divergence has often been used to measure prediction error or Bayesian surprise in the framework of predictive processing of the brain [6, 42, 43]. It is a metric used to measure the similarity between two different probability distributions. It represents how much information is lost when one probability distribution changes into another, and since it is non-negative, a small value indicates that the two distributions are similar. Specifically, it is calculated by taking the difference between the probability density functions of the two distributions, taking the logarithm at each point, and then computing the weighted average with respect to one of the distributions. The Kullback-Leibler divergence between two probability distributions P(x) and Q(x) is calculated using the following formula:

$D_{KL}(P||Q) = ΣP(i) \log (P(i)/Q(i))$ (3)

Here, P(i) and Q(i) represent the probabilities of selecting the value i according to the probability distributions P and Q, respectively. The duration of the temporal dynamics of surprise and uncertainty is dependent on the length of a song. To standardize this, we employed linear interpolation to match the entire duration of each song to that of the longest one in our dataset. Subsequently, the characteristics of the temporal dynamics of both surprise and uncertainty in each of pitch sequence, rhythm sequence, and pitch-rhythm sequence were dimensionally reduced to two dimensions using tSNE by python (random

state = 40, perplexity = 2, and early exaggenration = 20), allowing for visual representation of these statistical features.

## 5.3. Comparison of Acoustic Properties of Rhythm

All MIDI data were transformed into the WAV format. To ensure the sound intensity didn't affect the spectrotemporal modulation feature, the acoustic signals were first normalized using the z-score (mean = 0, SD = 1). We then analyzed these signals for their dominant amplitude modulation (AM) patterns below 40 Hz using the Bayesian probabilistic amplitude modulation model (PAD) [34]. It's known that acoustic signals comprise both slow-changing AM patterns and fast-changing carrier or frequency modulation (FM) patterns [29, 34]. The PAD model uses Bayesian inference to determine the modulators and carrier that best match the data and prior assumptions. This process infers the modulator and carrier from the signals based on set or learned parametric distributional constraints. The model's mathematical underpinnings involve specific likelihood functions and prior distributions based on signal samples and parameters θ. These parameters influence aspects like the typical modulator timescale or the carrier's frequency content.

In essence, PAD employs Bayesian inference to pinpoint the modulator and carrier that align best with the observed data and pre-existing beliefs. The end result is a probability distribution, which indicates how likely a given modulator and carrier setting is given the observed data. PAD then highlights the modulator and carrier with the highest probability, ensuring an optimal data fit. For our study, we manually set PAD parameters to obtain modulators below 40 Hz and carriers above 40 Hz. Carriers capture elements like noise and pitch. For every sample, these modulators were transformed into the time-frequency domain using scalograms, which represent the AM envelopes derived through iterative probabilistic amplitude demodulation. We then calculated and averaged the frequency power based on various musical factors like genre, instrument, and player.

We further explored patterns by determining the cycle lengths of the AM waveform. First, we identified troughs, which mark the boundaries between cycles. Having pinpointed all troughs, we assessed cycle lengths by measuring the distance between consecutive troughs. We then computed the horizontal rate. As an example, if one cycle is 500 ms and the next is also 500 ms, the horizontal rate is 0.5 (or 1:1) based on the formula $c_1/(c_1+c_2)$ [28]. These metrics were derived for each cycle in every musical sample, after which we averaged the cycle lengths for each piece of music.

# References

[1] Crossan, M. M., & Sorrenti, M. (2003). Making sense of improvisation. In Organizational improvisation (pp. 37-58). Routledge.

[2] McPherson, M. J., Lopez-Gonzalez, M., Rankin, S. K., & Limb, C. J. (2014). The role of emotion in musical improvisation: an analysis of structural features. PloS one, 9(8), e105144.

[3] Higgins, L., & Mantie, R. (2013). Improvisation as ability, culture, and experience. Music Educators Journal, 100(2), 38-44.

[4] Cowen, A. S., Elfenbein, H. A., Laukka, P., & Keltner, D. (2019). Mapping 24 emotions conveyed by brief human vocalization. American Psychologist, 74(6), 698.

[5] Khalil, R. A., Jones, E., Babar, M. I., Jan, T., Zafar, M. H., & Alhussain, T. (2019). Speech emotion recognition using deep learning techniques: A review. IEEE Access, 7, 117327-117345.

[6] Friston, K. J. (2010). The free-energy principle: A unified brain theory? Nature Reviews Neuroscience, 11(2), 127-138.

[7] Vuust, P., Heggli, O.A., Friston, K. J., & Kringelbach, M. L. (2022). Music in the brain. Nature Reviews Neuroscience, 23(5), 287-305.

[8] Koelsch, S., Vuust, P., & Friston, K. (2019). Predictive processes and the peculiar case of music. Trends in cognitive sciences, 23(1), 63-77.

[9] Daikoku, T. (2018). Neurophysiological markers of statistical learning in music and language: Hierarchy, entropy and uncertainty. Brain sciences, 8(6), 114.

[10] Pearce, M. T., & Wiggins, G. A. (2012). Auditory expectation: the information dynamics of music perception and cognition. Topics in cognitive science, 4(4), 625-652.

[11] Saffran, J. R. (2018). Statistical learning as a window into developmental disabilities. Journal of Neurodevelopmental Disorders, 10(35), December 2018.

[12] Daikoku, T., & Yumoto, M. (2023). Order of statistical learning depends on perceptive uncertainty. Current Research in Neurobiology, 4, 100080.

[13] Daikoku, T. (2018). Entropy, uncertainty, and the depth of implicit knowledge on musical creativity: computational study of improvisation in melody and rhythm. Frontiers in Computational Neuroscience, 12, 97.

[14] Zioga, I., Harrison, P. M., Pearce, M. T., Bhattacharya, J., & Luft, C. D. B. (2020). From learning to creativity: Identifying the behavioural and neural correlates of learning to predict human judgements of musical creativity. NeuroImage, 206, 116311.

[15] Daikoku, T., & Yumoto, M. (2020). Musical expertise facilitates statistical learning of rhythm and the perceptive uncertainty: A cross-cultural study. Neuropsychologia, 146, 107553.

[16] Przysinda, E., Zeng, T., Maves, K., Arkin, C., & Loui, P. (2017). Jazz musicians reveal role of expectancy in human creativity. Brain and cognition, 119, 45-53.

[17] Daikoku, T. (2018). Musical creativity and depth of implicit knowledge: spectral and temporal individualities in improvisation. Frontiers in computational neuroscience, 12, 89.

[18] Daikoku, T., Tanaka, M., & Yamawaki, S. (2023). Body Maps of Uncertainty and Surprise in Musical Chord Progression and its individual differences in Depression and Body Perception Sensitivity. bioRxiv, 2023-09.

[19] Cheung, V. K., Harrison, P. M., Meyer, L., Pearce, M. T., Haynes, J. D., & Koelsch, S. (2019). Uncertainty and surprise jointly predict musical pleasure and amygdala, hippocampus, and auditory cortex activity. Current Biology, 29(23), 4084-4092.
[20] Burkholder, J. P., Grout, D. J., & Palisca, C. V. (2019). A history of western music: Tenth international student edition. WW Norton & Company.
[21] Taruskin, R. (2005). The Oxford history of Western music. Oxford univ. press.
[22] Daikoku, T. (2019). Depth and the uncertainty of statistical knowledge on musical creativity fluctuate over a composer's lifetime. Frontiers in computational neuroscience, 13, 27.
[23] Daikoku, T., Kamermans, K., & Minatoya, M. (2023). Exploring cognitive individuality and the underlying creativity in statistical learning and phase entrainment. EXCLI journal, 22, 828.
[24] Altmann, G. (2017). Abstraction and generalisation in statistical learning: Implications for the relationship between semantic types and episodic tokens. Philosophical Transactions of the Royal Society B: Biological Sciences, 372(1711), 20160060.
[25] Daikoku, T., Wiggins, G. A., & Nagai, Y. (2021). Statistical Properties of Musical Creativity: Roles of Hierarchy and Uncertainty in Statistical Learning. Frontiers in Neuroscience, 15, 640412.
[26] Patel, A. D. (2003). Language, music, syntax and the brain. Nature neuroscience, 6(7), 674-681.
[27] Shannon, C. E. (1951). Prediction and entropy of printed english. Bell Syst. Tech. J. 30, 50–64. doi: 10.1002/j.1538-7305.1951.tb01366.x
[28] Roeske, T. C., Tchernichovski, O., Poeppel, D., & Jacoby, N. (2020). Categorical rhythms are shared between songbirds and humans. Current Biology, 30(18), 3544-3555.
[29] Daikoku, T., & Goswami, U. (2022). Hierarchical amplitude modulation structures and rhythm patterns: Comparing Western musical genres, song, and nature sounds to Babytalk. PloS One, 17(10), e0275631.
[30] Mehr, S. A., Singh, M., Knox, D., Ketter, D. M., Pickens-Jones, D., Atwood, S., ... & Glowacki, L. (2019). Universality and diversity in human song. Science, 366(6468), eaax0868.
[31] Daikoku, T., Yatomi, Y., & Yumoto, M. (2015). Statistical learning of music-and language-like sequences and tolerance for spectral shifts. Neurobiology of learning and memory, 118, 8-19.
[32] Daikoku, T., Yatomi, Y., & Yumoto, M. (2017). Statistical learning of an auditory sequence and reorganization of acquired knowledge: A time course of word segmentation and ordering. Neuropsychologia, 95, 1-10.
[33] Pfleiderer, M., Frieler, K., Abeber, J., Zaddach, W.-G., & Burkhard, B. (Eds.). (2017). Inside the Jazzomat. New perspectives for Jazz research. Schott Campus.
[34] R. E. Turner, and M. Sahani, "Demodulation as probabilistic inference," IEEE Transactions on Audio, Speech, and Language Processing, vol.19, no.8, pp.2398–2411, November 2011.
[35] Levitin, D. J., Chordia, P., & Menon, V. (2012). Musical rhythm spectra from Bach to Joplin obey a 1/f power law. Proceedings of the National Academy of Sciences, 109(10), 3716-3720.
[36] Schellenberg, E. G., Krysciak, A. M., & Campbell, R. J. (2000). Perceiving emotion in melody: Interactive effects of pitch and rhythm. Music Perception, 18(2), 155-171.

JAZZ IMPROVISATION AND ITS INDIVIDUALITY

[37] Banse, R., & Scherer, K. R. (1996). Acoustic profiles in vocal emotion expression. Journal of personality and social psychology, 70(3), 614.
[38] Tzanetakis, G., & Cook, P. (2002). Musical genre classification of audio signals. IEEE Transactions on speech and audio processing, 10(5), 293-302.
[39] Rutar, D., de Wolff, E., Kwisthout, J., & Hunnius, S. (2022). Statistical learning mechanisms are flexible and can adapt to structural input properties. Available at SSRN 4027230.
[40] Marko, M. K., Haith, A. M., Harran, M. D., & Shadmehr, R. (2012). Sensitivity to prediction error in reach adaptation. Journal of neurophysiology, 108(6), 1752-1763.
[41] Hasson, U. The neurobiology of uncertainty: Implications for statistical learning. Philos. Trans. R. Soc. Lond. B Biol. Sci. 2017, 372, 1711.
[42] Baldi P, Itti L. Of bits and wows: A Bayesian theory of surprise with applications to attention. Neural Netw. 2010;23:649–666. doi: 10.1016/j.neunet.2009.12.007.
[43] Itti L, Baldi P. Bayesian surprise attracts human attention. Vision Res. 2009;49:1295–1306. doi: 10.1016/j.visres.2008.09.007.